\documentclass[twocolumn,showpacs,preprintnumbers,amsmath,amssymb,prb]{revtex4}

\usepackage{graphicx}
\usepackage{dcolumn}
\usepackage{bm}

\begin{document}

\title{Spin-charge filtering through a spin-orbit coupled quantum dot
       controlled via an Aharonov-Bohm interferometer}

\author{R. J. Heary, J. E. Han, and Lingyin Zhu}
\affiliation{
Department of Physics, State University of New York at Buffalo, Buffalo, NY 14260, USA}

\date{\today}
\begin{abstract}
We show that a strongly correlated quantum dot embedded in an
Aharonov-Bohm interferometer can be used to filter both charge
and spin at zero voltage bias.  The magnitude with which the
Aharonov-Bohm arm is coupled to the system
controls the  many-body effects on
the quantum dot. When the quantum dot is in the Kondo regime
the flow of charge through the system can be tuned by
the phase of the Aharonov-Bohm arm, $\varphi_{AB}$.
Furthermore when  a spin-orbit interaction is present
on a Kondo quantum dot we can control the flow of spin by
the spin-orbit phase, $\varphi_{SO}$. The
existence of the Kondo peak at the Fermi energy makes it
possible to control the flow of both charge and spin in the
zero voltage bias limit.
\end{abstract}


\maketitle

\section{Introduction}

\begin{figure}[h]
\includegraphics[angle=0, width=6.0cm]{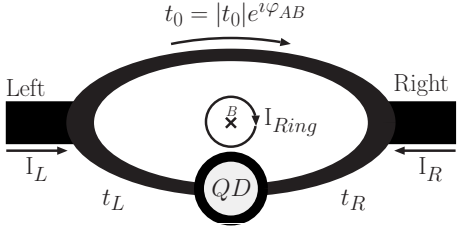}
\caption{
Experimental setup of the quantum dot embedded AB interferometer. Two infinite
electron reservoirs (Left, Right) are connected by two arms.
The bottom arm contains the embedded QD with real tunneling
coefficients $t_L$ and $t_R$.  The upper arm is a direct
connection between the left and right leads with a complex
tunneling coefficient $t_0$ = $|t_0|e^{i\varphi_{AB}}$, where
$\varphi_{AB}$ is a phase factor controlled by the magnetic
field, $B$.
The sign of $\varphi_{AB}$ is positive for electrons traveling
from the left to the right, as the arrow indicates.
$\mathrm{I}_L$, $\mathrm{I}_R$, and $\mathrm{I}_{Ring}$ are
the Left, Right, and Ring currents respectively.
}\label{fig:experiment}
\end{figure}

The ability to easily control charge and spin transport is
of great importance in nanotechnology, specifically
spintronics.  The Quantum Dot Aharonov-Bohm Interferometer
(QD-ABI) (Figure~\ref{fig:experiment}) has been found to be a
candidate for manipulating electron spins~\cite{spin polarization, Sun}.  In recent
years a number of theories~\cite{Kondo Fano, AB thermopower} have been put forth to take
advantage of the interference effects in such a geometry.
Sun and Xie~\cite{spin polarization} showed that the spin polarization
on the QD may be controlled via the voltage bias.  In the
presence of a local coulomb interaction on the QD Hofstetter
\textit{et al.}~\cite{Kondo Fano} studied the dependence of
the Fano line shape on the AB phase.

In this paper we show that the addition of a spin-orbit (SO)
interaction on the strongly correlated QD allows the QD-ABI
to function as a spin-charge filter.  The ability of this system to act as a filter is a
consequence of the interference between the continuum (AB arm)
and the localized state (QD) , which is known as the Fano
effect~\cite{U. Fano}.  Aside from the QD-ABI the Fano effect
has been observed in a single electron transistor~\cite{SET},
a quantum wire with a side coupled QD~\cite{QW-QD, QW-QD
Kondo}, and multiwall carbon nanotubes in a crossed
geometry~\cite{C nanotubes}.

The presence of the AB arm tends to localize electrons on
the QD, depending on how strongly the arm is coupled to the
system.  Therefore this property will give us an effective
control over the strength of these correlations on the QD. We
may exploit this tuning of the many-body physics to control the
charge-spin transport of the QD-ABI system.

The QD interactions we will study in this paper are an
on-site Coulomb interaction and the Rashba spin-orbit
interaction~\cite{Rashba}.
The Coulomb interaction in the
Kondo regime will allow us to filter both charge and spin at
zero bias.  The reason for this is due to the fact that the
Kondo effect induces a sharp resonance in the QD spectral
function at the Fermi energy.  The further addition of the
SO interaction, which is induced by the application of a gate
potential, for a single orbital QD, will create a spin
dependent phase factor~\cite{Sun} in the AB arm tunneling
coefficient.

Often the spin-orbit interaction is considered a coupling with
spin degrees of freedom mediated by interlevel transitions in
quantum dot systems and therefore, due to the significant
level spacing and the Coulomb interaction, the inter-level SO
coupling strengths in QD systems are thought to be small.
However, as pointed out in Ref.~\cite{Sun}, an intra-level phase
factor induced by the SO coupling may be realized in
high $g$-factor systems~\cite{Dobers} such as InGaAs quantum
dots.  Without the interference effect, {\it eg.} without the
AB-arm, such effect can be ignored. However, the presence of
the AB-arm not only manifests the intra-level SO interference
effects but also effectively controls the many-body effects on
the QD, hence a strong influence on the charge-spin transport.

The experimental setup of the QD-ABI is consistent with that
of Kobayashi \textit{et al.}~\cite{Fano}. The left and right
leads are modeled as infinite non-interacting electron
reservoirs. They are connected to each other via two arms.
The top arm is the AB arm which has a complex tunneling
coefficient $t_0 = |t_0|e^{i\varphi_{AB}}$ that is controlled by a magnetic field.
The sign of $\varphi_{AB}$ is positive for electrons traveling from the
left reservoir to the right reservoir.  We
consider the magnetic field to be small enough so that we may
ignore the Zeeman splitting of the QD energy level. The bottom
arm contains the embedded QD with real tunneling coefficients
$t_L$ and $t_R$.  We choose $t_L$, $t_R$ to be real because
the $L$, $R$ states are defined to absorb the phase factor. Our
calculations are carried out in the low-bias, linear response
regime.

\section{Theory}
\subsection{Transmission coefficient $T(\epsilon)$}

In this section of the paper we will derive the general transport
functions for the QD-ABI, which are correct with the
interaction on the QD. The non-interacting Hamiltonian of the system is given by
\begin{equation}
{\mathcal{H}} = {\mathcal{H}}^{ }_{L,R} + {\mathcal{H}}^{ }_d+
{\mathcal{H}}^{ }_t.
\end{equation}
The Hamiltonian $\mathcal{H}$ consists of three parts:  ${\mathcal{H}}_{L,R}$
describes the reservoirs, ${\mathcal{H}}_d$ the QD, and ${\mathcal{H}}_t$ the
tunneling between the reservoirs.
\begin{equation}
{\mathcal{H}}^{ }_{L,R} = \sum_{\alpha k\sigma} \epsilon^{ }_{\alpha k}
c^{\dagger}_{\alpha k\sigma}c^{ }_{\alpha k\sigma} \\
\end{equation}
\begin{equation}
{\mathcal{H}}^{ }_d = \sum_{\sigma}\epsilon^{ }_{d}d^{\dagger}_{\sigma}d^{ }_{\sigma} \\
\end{equation}
\begin{equation}
{\mathcal{H}}^{ }_t = -\frac{1}{\sqrt{\Omega}} \sum_{\alpha k\sigma}t^{ }_{\alpha}
(c^{\dagger}_{\alpha k\sigma}d^{ }_{\sigma} + h.c.) -
\frac{1}{\Omega} \sum_{k,k'} (t^{ }_0 c^{\dagger}_{L k\sigma} c^{ }_{R k'\sigma} + h.c.) ,
\end{equation}
where $c^\dag_{\alpha k\sigma}(c_{\alpha k\sigma})$ and
$d^\dag_{\sigma}(d_{\sigma})$ are the creation (annihilation) operators
with momentum $k$ and spin $\sigma$ of the $\alpha = (L,R)$ reservoir and the
QD, respectively.  In addition, $\Omega$ is the volume of the reservoirs which
is taken to infinity.

Before presenting the Landauer formula for the current
we define the parameters.  In the non-interacting limit without the AB arm,
the line broadening of the QD spectral function due to the leads is
$\Gamma = \Gamma_L+\Gamma_R$ where $\Gamma_\alpha = \pi N_0t_\alpha^2$.  In
our calculations we take the density of states, $N_0$,  to be a constant.

The exact current from the $L$ to $R$ reservoir, regardless of the local
interaction on the QD, is given by the
Landauer formula~\cite{Wingreen Meir},
\begin{equation}
I_L = \frac{2e^2}{h}\int_{-\infty}^{\infty}
     T(\epsilon)\Delta f(\epsilon)d\epsilon .
\end{equation}
Here $\Delta f(\epsilon) = f_L(\epsilon)-f_R(\epsilon)$ and
$T(\epsilon)$ is the transmission function.  The transmission
function was previously reported in Refs.~\cite{Kondo Fano, AB
thermopower}, although without derivation.  Therefore we present
the derivation, which makes use of standard Keldysh Green function
techniques~\cite{Keldysh1, Keldysh2}, in the Appendix. Here we
summarize the results.

The transmission function may first be decoupled into two parts, the flow
of current through the QD and through the AB arm,
\begin{eqnarray}\label{T}
T(\epsilon)\Delta f &=& \mathrm{i^{ }_{QD}}(\epsilon) + \mathrm{i^{ }_{AB}}(\epsilon) \\
\mathrm{i^{ }_{QD}}(\epsilon) &=& -t^{ }_{L}(G^<_{dL}(\epsilon) - G^<_{Ld}(\epsilon))\nonumber
                                  = 2\mathrm{Re}[-t^{ }_{L}G^<_{dL}(\epsilon)]\\
\mathrm{i^{ }_{AB}}(\epsilon) &=& \left[-t^*_0G^<_{RL}(\epsilon) + t^{ }_0 G^<_{LR}(\epsilon)\right]\nonumber
                                  = 2\mathrm{Re}[-t^*_{0}G^<_{RL}(\epsilon)],
\end{eqnarray}
where we have used the relation, $G^<_{\alpha\beta} = -(G^<_{\beta\alpha})^*$.
$\mathrm{i^{ }_{QD}}$ and $\mathrm{i^{ }_{AB}}$ are the contributions
to the current from the QD and the AB arm respectively.
To simplify our notation we define the following parameters
\begin{eqnarray}
T^{ }_0 &=& \frac{4r^{ }_0}{(1+r^{ }_0)^2} \\
R^{ }_0 &=& 1-T^{ }_0 = \left[\frac{1-r^{ }_0}{1+r^{ }_0}\right]^2 \\
\alpha &=& \frac{4\Gamma^{ }_L\Gamma^{ }_R}{\Gamma^2} \\
\label{Gammaprime}
\bar{\Gamma} &=& \frac{\Gamma}{1+r^{ }_0},
\end{eqnarray}
where $r_0 = \pi^2N_0^2|t_0|^2$.
Here $T_0$ is the transmission function when the QD is disconnected from the
left and right reservoirs, i.e. $t_L=t_R=0$.  The current through the QD and AB arm is found below.
\begin{widetext}
\begin{eqnarray}\label{i_QD}
\mathrm{i^{ }_{QD}} = & & \left[-\alpha\bar{\Gamma}\sqrt{R^{ }_0}
               -2T^{ }_0\Gamma^{ }_L\Gamma^{ }_R\sin^2\left(\varphi^{ }_{AB}\right)
               -\frac{\Gamma_L-\Gamma_R}{2}\sqrt{\alpha T^{ }_0}\sin\left(\varphi^{ }_{AB}\right)\right]\mathrm{Im}[ G^R_{dd}]\Delta f - \nonumber \\
         & &   \bar{\Gamma}\sqrt{\alpha T^{ }_0}\cos\left(\varphi^{ }_{AB}\right)
               \mathrm{Re}[G^R_{dd}]\Delta f  - \mathrm{i^{ }_{Ring}}  \\
\mathrm{i^{ }_{AB}} = & & \left[\alpha\bar{\Gamma}\sqrt{T^{ }_0} \left(
               \sqrt{T^{ }_0}\cos^2\left(\varphi^{ }_{AB}\right) - 1\right)
               + 2T^{ }_0\Gamma^{ }_L\Gamma^{ }_R\sin^2\left(\varphi^{ }_{AB}\right)
               +\frac{\Gamma_L-\Gamma_R}{2}\sqrt{\alpha T^{ }_0}\sin\left(\varphi^{ }_{AB}\right)\right]\mathrm{Im}[ G^R_{dd}]\Delta f + \nonumber \\
        & &    2T^{ }_0\bar{\Gamma}\sqrt{\alpha}
               \cos\left(\varphi^{ }_{AB}\right) \mathrm{Re}[ G^R_{dd}]\Delta f  +
               \mathrm{i^{ }_{Ring}} + T^{ }_0\Delta f.
\label{i_AB}
\end{eqnarray}
\end{widetext}

In our analysis we find there exists a ring current even at
zero bias,
\begin{equation}
\mathrm{i^{ }_{Ring}} = -\sqrt{\alpha T^{ }_0}\Gamma\sin\varphi^{ }_{AB}
\mathrm{Im}[ G^R_{dd}]\bar{f} .
\end{equation}
Here $\bar{f} = (f_L+f_R)/2$ is the average Fermi function.
This current flows in the clockwise direction through the AB
ring and persists even at zero bias since it is proportional
to $\bar{f}$, not $\Delta f$.
Although the ring current can be of the same order of magnitude
as the total current, it does not contribute to the
source-drain current.

Using Eq.~(\ref{T}) we arrive at the exact transmission function,
\begin{equation}\label{Transmission}
\begin{split}
T(\epsilon) &=
                 T^{ }_0 - 2\bar{\Gamma}\sqrt{\alpha T^{ }_0R^{ }_0}\cos(\varphi^{ }_{AB})            \mathrm{Re}[G^R_{dd}(\epsilon)] \\
              & - \bar{\Gamma}[\alpha(1-T^{ }_0 \cos^2(\varphi_{AB}))-T^{ }_0] \mathrm{Im}[ G^R_{dd}(\epsilon)] .
\end{split}
\end{equation}
We emphasize that $G^R_{dd}(\epsilon)$ is the full interacting
QD retarded Green function, and Eq.~\ref{Transmission} applies
to systems with an interacting QD.

\subsection{Non-interacting limit}

In the non-interacting limit we will address two important
points. First, as the coupling to the left and right reservoirs is
increased through $|t_0|$, the electron becomes more localized
on the QD.  Secondly, the noninteracting system is not suitable for
controlling both charge and spin transport.

\begin{figure}[h]
\includegraphics[angle=270, width=8.0cm]{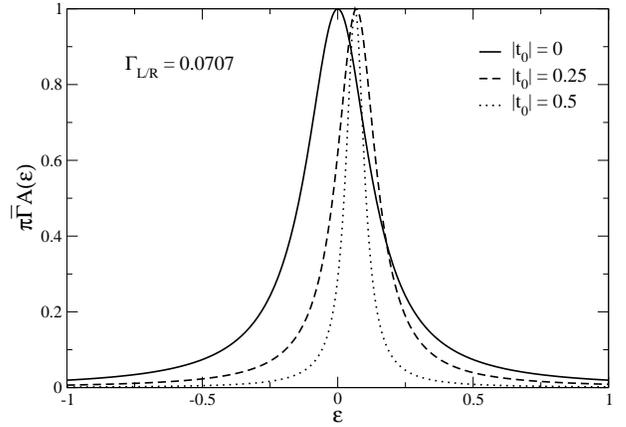}
\caption{QD spectral function for different magnitudes of
$|t_0|$ at $\varphi_{AB} = 0$.
As $|t_0|$ is increased the electron becomes more localized
on the QD and the distribution of energies shift away from
the Fermi energy.}
\label{fig:nonint_spectral}
\end{figure}

The noninteracting QD spectral function is given by
\begin{equation}
A(\epsilon) = \frac{\bar{\Gamma}/\pi}{(\epsilon-\epsilon^{ }_d - \delta)^2 +\bar{\Gamma}^2} ,
\end{equation}
where
\begin{equation}
\label{delta}
\delta = \frac{2\sqrt{r^{ }_0\Gamma^{ }_L\Gamma^{ }_R}}{1+r^{ }_0}\cos(\varphi^{ }_{AB}) .
\end{equation}
and $\bar{\Gamma}$ was defined in Eq.~(\ref{Gammaprime}).
We plot $A(\epsilon)$ in Figure~\ref{fig:nonint_spectral} for
different values of $|t_0|$.  Notice that as the magnitude
of $|t_0|$ is increased, the QD spectral function becomes
sharper and the center of the peak is shifted.
This shift in the QD energy, $\delta$ and the reduced
line broadening, $\bar{\Gamma}$ are easily understood by
performing a bonding-anti-bonding transformation of the leads in
real space.  Upon doing so the QD is now only coupled to the
first site of the bonding chain where the local
energy of this site is shifted by $-|t_0|\cos(\varphi_{AB})$.
Thus by increasing $|t_0|$, the connection of the QD to the
bonding chain is reduced, and the energy level of the QD is shifted upward.

\begin{figure}[h]
\includegraphics[angle=270, width=8.0cm]{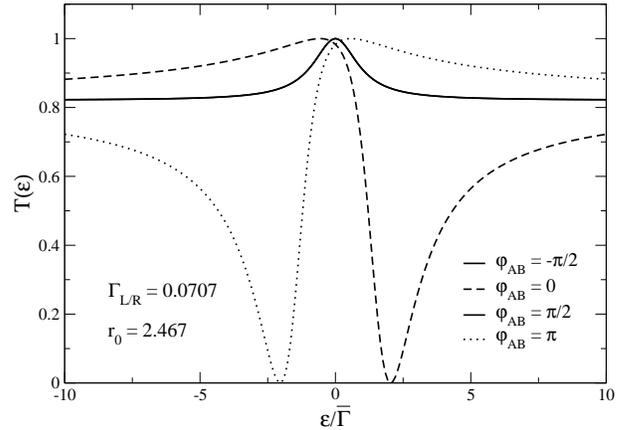}
\caption{
Noninteracting transmission function for different values of
$\varphi_{AB}$.
When $\varphi_{AB}$ = $\left(-\frac{\pi}{2},
\frac{\pi}{2}\right)$ we see a resonance at the QD energy
level.  The $\varphi_{AB}$ = $\left(-\frac{\pi}{2},
\frac{\pi}{2}\right)$ curves coincide, away from these curves
the interference effects become more
pronounced and the transmission function becomes asymmetrical.
When $\varphi_{AB}$ = $\left(0, \pi\right)$ the interference
effects become most pronounced and a very strong anti-resonance
emerges which is shifted to the left or right of the QD energy
level. In the limit $|\epsilon|\rightarrow \infty$ the
transmission converges to a finite value, due to the AB arm.}
\label{fig:nonint_transmission}
\end{figure}

The differential conductance is given as $G(\Phi)=e(dI/d\Phi)$,
and in the zero bias and low-temperature limit $G(0) =
\frac{2e^2}{h}T(0)$.  Therefore at equilibrium we only need
$T(0)$ to determine the conductance.  In
Figure~\ref{fig:nonint_transmission} we plot the transmission
amplitude as a function of the AB phase, $\varphi_{AB}$.  When
$\varphi_{AB}=\pm\frac{\pi}{2}$, $T(0)=1$.  When
$\varphi_{AB}= (0,\pi)$, the Fano anti-resonance becomes most
prominent and the minimum of the peaks are positioned significantly
above and below the Fermi energy.  As a result of this fact
we are not able to extinguish the charge or spin conductance at
zero bias.  If we were able to shift
the minima of these anti-resonance peaks to the Fermi energy,
then we would be able to fully control the transport through
this system at zero bias by tuning $\varphi_{AB}$. In the
next section of this paper we show that we can accomplish this
through the Kondo effect.

\subsection{Interacting Fano Effect}

\begin{figure}[h]
\includegraphics[angle=0, width=6.0cm]{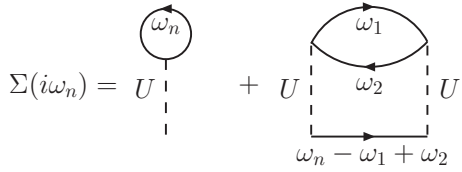}
\caption{Self energy expanded to second order in U.}
\label{fig:self_diagram}
\end{figure}

When a local many-body interaction is present on the QD site,
the problem is essentially reduced to calculating the full QD
retarded Green function,  $G^R_{dd}(\epsilon)$.  Once we know
this Green function we may then calculate the transmission
function using Eq. (\ref{Transmission}).  In this paper we
take into account the Coulomb interaction on the QD in the
form of the Anderson interaction,
\begin{equation}
{\mathcal{H}_{int}} = U\hat{n}_{d\uparrow}\hat{n}_{d\downarrow} .
\end{equation}
We perform the diagrammatic calculation in the imaginary-time formalism.
The QD Green function, at imaginary Matsubara frequency
$i\omega_n = i\frac{(2n+1)\pi}{\beta}$, is given by $G_{dd}(i\omega_n) =
[(G_{dd}^0)^{-1}(i\omega_n)-\Sigma(i\omega_n)]^{-1}$,
where the self energy, $\Sigma(i\omega_n)$ is calculated to second-order
in $U$ (Figure~\ref{fig:self_diagram}).  The second-order
perturbation theory has been studied
extensively~\cite{U2_1,U2_2,U2_3,U2_4,U2_5} and shown to be a very good
approximation in the particle-hole symmetric limit up to values of
$U/\bar{\Gamma}\simeq 6$ (weak-coupling regime).
This weak-coupling approximation has also been used within the framework
of dynamical mean field theory~\cite{DMFT}, and produced agreement with
nonperturbative methods.  In our model the particle-hole symmetry will be broken for
$\delta\neq 0$, but the calculation of the zero bias conductance is
in excellent agreement with the nonperturbative numerical
renormalization group~\cite{Kondo Fano}(NRG) as will be shown.

The first-order diagram becomes $U\langle n_d\rangle = \frac{U}{2}$ in
the half-filled limit.  Absorbing the first-order diagram into the
non-interacting Matsubara Green function we have
\begin{equation}
G_{dd}^0(i\omega_n) = \frac{1}{i\omega_n -\delta +i\bar{\Gamma}
\cdot\mathrm{Sign}(\omega_n)} ,
\end{equation}
and the second order self energy becomes
\begin{equation}
\Sigma^{(2)}(i\omega_n) = \frac{U^2}{\beta^2}
       \sum_{\omega_1,\omega_2} G_{dd}^0(i\omega_n-i\omega_1)
       G_{dd}^0(i\omega_2)G_{dd}^0(i\omega_1+i\omega_2) .
\end{equation}

Now we are in position to solve this problem numerically.
First we calculate the self-energy in Matsubara frequency.
Then in order to calculate $T(\epsilon)$ we must analytically
continue the Green function to its retarded form in real
frequency space, $G_{dd}(i\omega_n\rightarrow\epsilon
+i\eta)$. For the numerical analytical continuation we used
the $N$-Point Pad$\acute{\mathrm{e}}$ approximant
method~\cite{pade}.

\begin{figure}[h]
\includegraphics[angle=270, width=8.0cm]{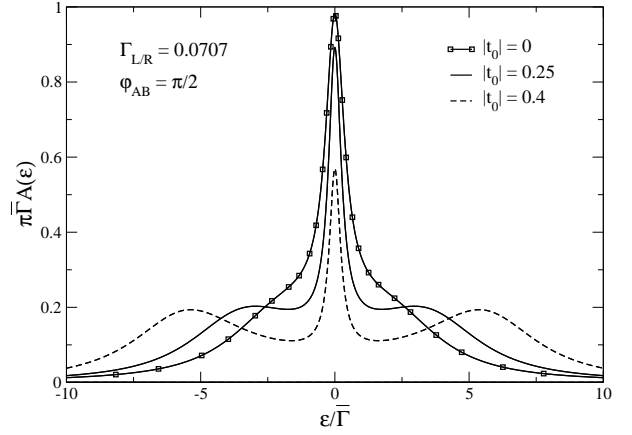}
\caption{
Interacting QD spectral function for different magnitudes of
$|t_0|$ when $\varphi_{AB} = \frac{\pi}{2}$, U = 0.6 and
$\beta$ = 160.  We see that as $|t_0|$ is increased the
electron becomes more localized on the QD and as a result the
correlations due to the Coulomb interaction become more
pronounced. }\label{fig:int_spectral t0}
\end{figure}

Let us first look at the spectral function, $A(\epsilon) =
-\frac{1}{\pi} {\mathrm{Im}}[G_{dd}(\epsilon)]$.  For the
resonant case ($\varphi_{AB} = \frac{\pi}{2}$) the spectral
function is plotted for different values of $|t_0|$ in
Figure~\ref{fig:int_spectral t0}.  We see that as we increase
the coupling $|t_0|$ of the L, R states to the AB arm, the effective many-body
interaction $U/\bar{\Gamma}$ is strongly enhanced.  From our analysis of the
noninteracting system this is exactly what we expected to
happen.  When $|t_0| = 0$ we start in the weakly interacting valence
fluctuating regime, but as we increase $|t_0|$ to 0.4 the Kondo
at zero frequency and the Hubbard satellites~\cite{spectral} emerge.

\begin{figure}[h]
\includegraphics[angle=270, width=8.0cm]{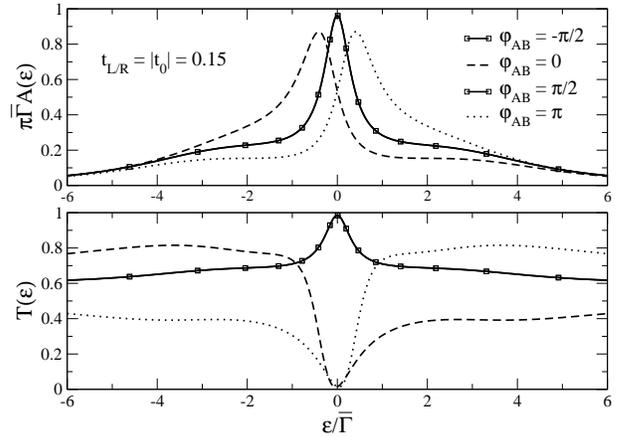}
\caption{
QD spectral function and transmission amplitude for U = 0.6
and $\beta$ = 160.  The $\varphi_{AB} = \frac{\pi}{2},
\frac{-\pi}{2}$ curves are identical, and at these
values the spectral function is symmetric and the
transmission amplitude displays resonant behavior.  On the
other hand when $\varphi_{AB} = 0, \pi$ the spectral function
is asymmetric and the transmission amplitude has an
anti-resonance.  }\label{fig:int_spectral trans}
\end{figure}

\begin{figure}[h]
\includegraphics[angle=270, width=8.0cm]{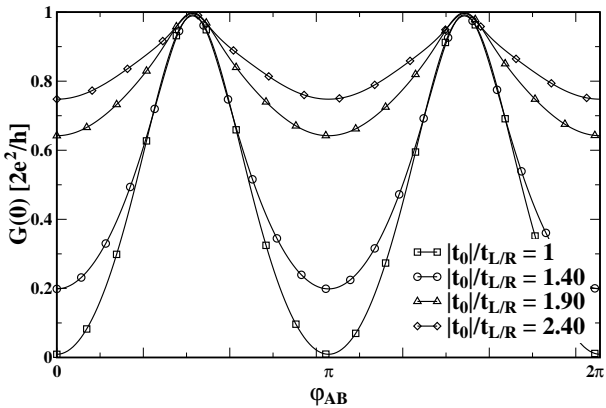}
\caption{Zero bias conductance as a function of $\varphi_{AB}$
for $t_L=t_R=0.15$, U = 0.5 and $\beta$ = 160.  The conductance approaches
unity when $\varphi_{AB} = \frac{2n+1}{2}\pi$.  Conversely the
conductance approaches a minimum when $\varphi_{AB} = n\pi$, with
integer $n$.  }\label{fig:AB_oscillations}
\end{figure}

The transmission functions is given in
Figure~\ref{fig:int_spectral trans}.  As in the noninteracting
case we have resonance (anti-resonance) phenomena at
$\varphi_{AB} = -\frac{\pi}{2}, \frac{\pi}{2}$($0, \pi$)
respectively.  The anti-resonance peaks are known as the
Fano-Kondo anti-resonance and have been observed experimentally
in a quantum wire with a side coupled QD~\cite{QW-QD, QW-QD
Kondo}.  For $\varphi_{AB} = 0, \pi$ the anti-resonance peaks
are antisymmetric of one another about the Fermi energy.

Now let us examine the zero bias conductance.  Our results are
given in Figure~\ref{fig:AB_oscillations} as a function of
$\varphi_{AB}$ for different values of the ratio $|t_0|/t_{L/R}$.
The NRG calculation of
Ref.~\cite{Kondo Fano} examined the zero bias conductance
as a function of $\varphi_{AB}$ for different values of the gate
potential, $\epsilon_d$.   A careful straightforward identification
of the model parameters shows that our results agree
excellently with the NRG~\cite{comparison},
which justifies our self-energy approximation.  Since $G(0)$ is
only dependent upon the value of the Green function at the Fermi
energy, this implies that the second-order self energy approximation
at least produces reliable results near the Fermi energy for
$\delta\neq 0$.

From Figure~\ref{fig:AB_oscillations} we see that $G(0)$ oscillates
as a function of $\varphi_{AB}$ and the magnitude of these
oscillations is strongly dependant on the ratio,
$|t_0|/t_{L/R}$.  We find that $|t_0|/t_{L/R} = 1$  is a special case which
maximizes the magnitude of the AB oscillations.
These oscillations are a consequence of the
$\cos^2(\varphi_{AB})$ term in Eq.~(\ref{Transmission})
and the sharp feature near the Fermi energy of the
spectral function (Figure~\ref{fig:int_spectral trans}).
The resonance ($G(0)=\mathrm{max}$)
[anti-resonance($G(0)=\mathrm{min}$)] peaks occur at
$\varphi_{AB} = \frac{2n+1}{2}\pi[n\pi]$, with integer $n$.
Thus by tuning $\varphi_{AB}$ we may filter the charge through the
system at zero voltage bias.  The reason why we have so much control over the
conductance at zero bias is due to the Kondo effect.  The
Kondo effect induces a sharp peak near the Fermi energy in the
spectral function, insensitive to $\varphi_{AB}$.  This feature
is due to the many body effect and is not present in
the non-interacting case (Figure~\ref{fig:nonint_transmission}).

\subsection{Spin Transport}

For a single orbital QD the addition of the spin-orbit interaction
induces a spin dependent phase in
$t_R$ ~\cite{Sun}, i.e. $t_R\rightarrow t_Re^{-i\sigma\varphi_{SO}}$ , where
$\sigma$ =+(-) for spin up(down).
Unlike $\varphi_{AB}$ which arises from the orbital motion,
$\varphi_{SO}$ depends on the spin.
To simplify our Hamiltonian we make the unitary transformation
\begin{equation}
c^\dag_{Rk}\rightarrow e^{is\varphi_{SO}}c^\dag_{Rk}
\end{equation}
so that
\begin{equation}
t_0 \rightarrow e^{is\varphi_{SO}}t_0
\end{equation}
Now we define a spin dependent AB tunneling coefficient
\begin{equation}
t_{0\sigma}= |t_0|e^{i(\varphi_{AB} +\sigma\varphi_{SO})} .
\end{equation}
As a result of the spin dependence in $t_{0\sigma}$, the
Green's functions also gain a spin dependence and
the spin dependant second order self energy is given
by
\begin{equation}
\begin{split}
\Sigma^{(2)}_\sigma(i\omega_n) = &\frac{U^2}{\beta^2}
       \sum_{\omega_1,\omega_2} G_{dd\sigma}^0(i\omega_n-i\omega_1) \times \\
      &G_{dd-\sigma}^0(i\omega_2)G_{dd-\sigma}^0(i\omega_1+i\omega_2)
\end{split}
\end{equation}

\begin{figure}[h]
\includegraphics[angle=270, width=8.0cm]{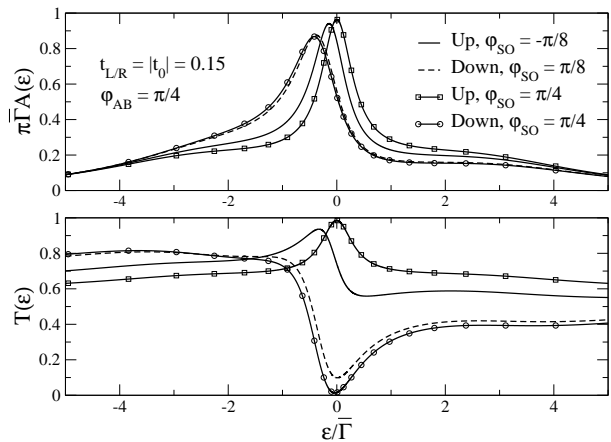}
\caption{Spin dependant spectral and transmission functions for
$\varphi_{SO}=\frac{\pi}{8}, \frac{\pi}{4}$, U = 0.6 and $\beta$ = 160.
For $\varphi_{SO}=\frac{\pi}{4}$ $A_\uparrow$($A_\downarrow$) is
symmetric(asymmetric), while $T_\uparrow$($T_\downarrow$) displays
resonance(anti-resonance) behavior.  In addition, $T_\uparrow(0)\sim 1$
and $T_\downarrow(0)\sim 0$.
For $\varphi_{SO}=\frac{\pi}{8}$ both $A_\uparrow$ and $A_\downarrow$ are
asymmetrical and $A_\downarrow$ lies on the curve of
$A_\downarrow(\varphi_{SO}=\frac{\pi}{4})$. $T_\uparrow$ and
$T_\downarrow$ both display anti-resonant phenomena.
}\label{fig:Spin1}
\end{figure}

Now let us examine the spectral and transmission functions given in
Figure~\ref{fig:Spin1}. As in the spin independent case we see
resonance and antiresonance behavior in the transmission function.
More importantly, due to the SO interaction we now
have two phase factors, $\varphi_{AB}$ and $\varphi_{SO}$,
which along with the Kondo resonance near the Fermi energy, we may use to
filter the spin up and spin down electrons independently.

Let us take a closer look at Figure~\ref{fig:Spin1}.  We see that when
$\varphi_{AB} = \varphi_{SO} = \frac{\pi}{4}$, $A_{\uparrow}$
has a spectral structure similar to that of a Kondo dot while $A_{\downarrow}$ develops
a slight asymmetry from the spin up case.  Looking at the transmission function, $T_{\uparrow}$
is similar in shape to its spectral function; and $T_{\uparrow}$ shows
strong antiresonance behavior due to the AB ring.  This mechanism gives us a strong control on the
spin-transport.  At the Fermi energy $T_{\uparrow}(0)\cong 1$ and $T_{\downarrow}\cong 0$.
When $\varphi_{SO} = \frac{\pi}{8}$ both $T_{\uparrow}$ and $T_{\downarrow}$
show strong interference effects.

\begin{figure}[h]
\includegraphics[angle=270, width=8.0cm]{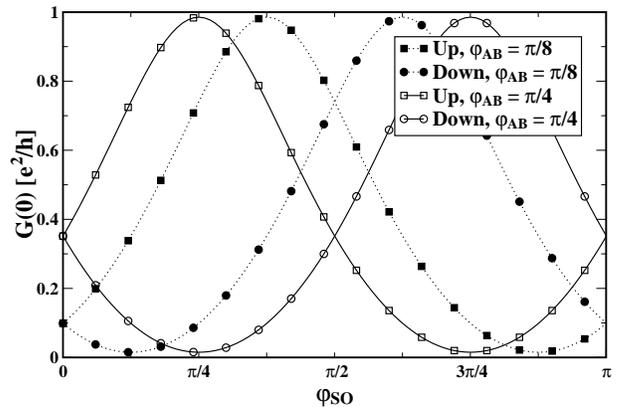}
\caption{SO oscillations for $\varphi_{AB}=\frac{\pi}{8}, \frac{\pi}{4}$,
$U$ = 0.6, and $\beta$ = 160.
The minima and maxima of the SO oscillations
approach 0 and 1 respectively.
}\label{fig:Spin2}
\end{figure}

To look at the zero bias spin conductance we turn to Figure~\ref{fig:Spin2} where
we show the SO oscillations of the zero bias conductance.  The behavior of the
spin up and spin down conductance is a result of
the $\cos^2(\varphi_{AB}+\varphi_{SO})$ term for spin up and
the  $\cos^2(\varphi_{AB}-\varphi_{SO})$ term for spin down in Eq.~(\ref{Transmission}).
In the figure we see that when $\varphi_{AB}  = \frac{\pi}{4}$ the spin polarized
conductance [$\eta = (G_\uparrow(0)-G_\downarrow(0))/(G_\uparrow(0)+G_\downarrow(0))$] is maximized.
Conversely when $\varphi_{SO} = \frac{\pi}{2}$ the spin polarization is suppressed.

\begin{figure}[h]
\includegraphics[angle=270, width=8.0cm]{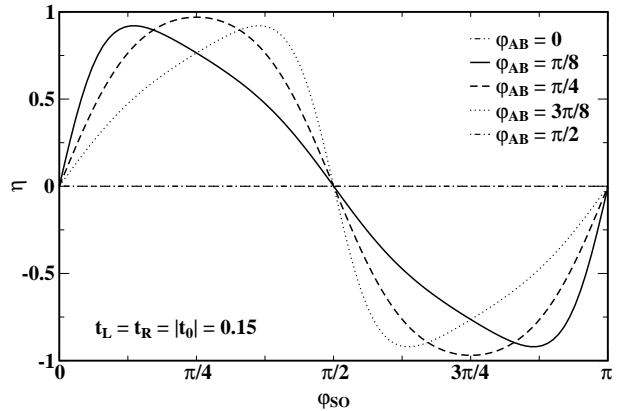}
\caption{Spin-polarized conductance for different AB phases when
U = 0.6 and $\beta$ = 160.  Here the
spin polarization is defined as, $\eta = [G_\uparrow(0)-G_\downarrow(0)]/
[G_\uparrow(0)+G_\downarrow(0)]$.  We see that the spin polarization
may be fully controlled by tuning $\varphi_{AB}$ and $\varphi_{SO}$. 
The $\varphi_{AB} = 0, \frac{\pi}{2}$ curves are identical.}\label{fig:spin_pol}
\end{figure}

The most important results of this paper are shown in
the spin polarized conductance in Figure~\ref{fig:spin_pol}.
Here we plot $\eta$ as a function of
$\varphi_{SO}$ for different values of $\varphi_{AB}$.
The maximum in the spin up/down conductance occurs approximately when
$\varphi_{AB} \pm \varphi_{SO}=\frac{2n+1}{2}\pi$, while the
minimum occurs at $\varphi_{AB} \pm \varphi_{SO}=m\pi$.
This can be seen from the $\cos^2(\varphi)$ factor in
Eq.~(\ref{Transmission}), since at the Kondo resonance ($\epsilon\approx 0$)
only $T_0$ and the term proportional to Im$[G^R_{dd}(\epsilon)]$
become relevant.  Therefore the
necessary conditions on $\varphi_{SO}$ and $\varphi_{AB}$ for the
spin polarized conductance to be maximized and minimized are
\begin{eqnarray}
\left(\frac{\varphi_{SO}}{\varphi_{AB}}\right)_{\eta \approx 1} &=&
                          \frac{n-m+\frac{1}{2}}{{n+m+\frac{1}{2}}} \\
\left(\frac{\varphi_{SO}}{\varphi_{AB}}\right)_{\eta \approx 0} &=& \frac{n-m}{n+m+1} .
\end{eqnarray}

Here again we emphasize the crucial role which the many-body
interactions play in this system.  The Kondo effect serves to
simultaneously pin the resonance/antiresonance peaks of the spin
dependent conductance at the chemical potential, as seen in
Figure~\ref{fig:int_spectral trans}. As a result, when the SO
interaction is turned on, we gain complete control over the spin
polarization (Figure~\ref{fig:spin_pol}).  Furthermore, even if there
only exists a small SO interaction in the QD, this device serves to
enhance those inherent SO effects.  In contrast to this, in the
noninteracting device (Figure~\ref{fig:nonint_transmission}) the
resonance/antiresonance peaks occur at significantly different energies,
making it impractical to use as a spin/charge filter.  Therefore we have
shown that the QD-ABI is an ideal spin filter where neither a voltage
bias or a gate potential is needed.

\section{Conclusion}

In this paper we have shown that the QD-ABI may
be used to localize and delocalize electrons on the QD.  As a
result, when the QD is strongly correlated we can effectively
control the strength of these correlations through the
magnitude of $t_0$. With larger $t_0$, electrons get more
localized on the QD.
In the case of a coulomb interaction on the QD, the Kondo
effect induces a sharp peak in the spectral function near the
Fermi energy.
We emphasize that the exact relation for the transmission,
Eq.~(\ref{Transmission}),
shows that the resonance/anti-resonance is driven by the AB
phase factor once the spectral function develops a sharp
feature near the Fermi energy, independent of its many-body
character.

The most important result of this paper is realized when a SO
interaction is present on the QD in addition to the coulomb
interaction.  In this case the Kondo peak and the spin dependent
$\cos^2(\varphi)$ term  induce the
resonance/anti-resonance in the zero bias spin dependant
conductance. Due to this property we can
vary the $G_\uparrow(0)$ and $G_\downarrow(0)$ between 0 and
$e^2/h$.  Further, the SO phase gives us another degree of
freedom, in addition to the AB phase, so that the spin up and
spin down conductances may be varied independently.  Thus
by tuning $\varphi_{AB}$ and $\varphi_{SO}$ the spin
polarization may be fully controlled in the QD-ABI.

\appendix
\section{Derivation of T($\epsilon$)}

In this appendix we present the full derivation of the transmission function.  The
transmission is given in terms of Keldysh Green functions,
\begin{eqnarray}
T(\epsilon)\Delta f &=& \mathrm{i^{ }_{QD}}(\epsilon) +
\mathrm{i^{ }_{AB}}(\epsilon)\\
\mathrm{i^{ }_{QD}}(\epsilon) &=& -t^{ }_{L}[G^<_{dL}(\epsilon)
- G^<_{Ld}(\epsilon)] \\
\mathrm{i^{ }_{AB}}(\epsilon) &=& \left[-t^*_0G^<_{RL}(\epsilon) +
t^{ }_0 G^<_{LR}(\epsilon)\right] ,
\end{eqnarray}
where $G^<_{dL}$ and $G^<_{RL}$ are the Fourier transforms of the
following time dependent nonequilibrium Green functions
(NEGFs),
\begin{eqnarray}
G^<_{dL}(t) &=& \frac{1}{\sqrt{\Omega}}\sum_k i\langle d^\dag_{ }(0) c^{ }_{Lk}(t)\rangle \\
G^<_{RL}(t) &=& \frac{1}{\Omega}\sum_{kk'} i\langle c^\dag_{Lk}(0) c^{ }_{Lk'}(t)\rangle  ,
\end{eqnarray}
and $G^<_{Ld}=-(G^<_{dL})^*$ and $G^<_{LR}=-(G^<_{RL})^*$.
In calculating these Green functions it becomes convenient to
disconnect the QD from the reservoirs, i.e. $t_L=t_R=0$, and define
the following QD-excluded Green functions
\begin{eqnarray}
g^r_{LL} &=& g^r_{RR} = \frac{-i\pi N^{ }_0}{1+r^{ }_0} \\
g^r_{RL} &=& \left(g^r_{LR}\right)^* = \frac{\pi^2N_0^2t^{ }_0}{1+r^{ }_0} \\
g^<_{LL} &=& \frac{2\pi iN_0}{\left(1+r^{ }_0\right)^2}\left(f^{ }_L+r^{ }_0f^{ }_R\right) \\
g^<_{RR} &=& \frac{2\pi iN_0}{\left(1+r^{ }_0\right)^2}\left(f^{ }_R+r^{ }_0f^{ }_L\right) \\
g^<_{RL} &=& -\left(g^<_{LR}\right)^* = \frac{-2\pi^2N_0^2}{\left(1+r_0^{ }\right)^2}
                                        t_0\left(f^{ }_L-f^{ }_R\right)  .
\end{eqnarray}
We express $G^<_{dL}$ and $G^<_{RL}$ in terms
of the fully interacting retarded and advance QD Green
functions and the above QD-excluded noninteracting NEGFs.

To simplify our notation we define $F^r_{\alpha d}$,
$F^a_{\alpha d}$, and $F^<_{\alpha d}$ which are the retarded,
advanced, and lesser Green functions which describe the
transport from the QD to the $\alpha$ lead in terms of the
QD-excluded Green functions.
\begin{widetext}
\begin{eqnarray}
F^r_{Ld} &=& \left(F^a_{dL}\right)^* = t^{ }_L\left(t^{ }_Lg^{ }_{LL}+t^{ }_Rg^{ }_{LR}\right)^r
             = \frac{1}{1+r^{ }_0}\left(-i\Gamma^{ }_L+\sqrt{r^{ }_0\Gamma^{ }_L\Gamma^{ }_R}
               e^{-i\varphi_{AB}}\right) \\
F^r_{Rd} &=& \left(F^a_{dR}\right)^* = t^{ }_R\left(t^{ }_Rg^{ }_{RR}+t^{ }_Lg^{ }_{RL}\right)^r
             = \frac{1}{1+r^{ }_0}\left(-i\Gamma^{ }_R+\sqrt{r^{ }_0\Gamma^{ }_L\Gamma^{ }_R}
               e^{i\varphi_{AB}}\right) \\
F^<_{Ld} &=& -(F^<_{dL})^*=
             t^{ }_L\left(t^{ }_Lg^{ }_{LL}+t^{ }_Rg^{}_{LR}\right)^<
             = \frac{2}{1+r^{ }_0}\left[i\Gamma^{ }_L\left(f^{ }_L+r^{ }_0f^{ }_R\right)+
             \sqrt{r^{ }_0\Gamma^{ }_L\Gamma^{ }_R} e^{-i\varphi_{AB}}
             \left(f^{ }_L-f^{ }_R\right)\right] \\
F^<_{Rd} &=& -(F^<_{dR})^* =
             t^{ }_R\left(t^{ }_Rg^{ }_{RR}+t^{ }_Lg^{}_{RL}\right)^<
             = \frac{2}{1+r^{ }_0}\left[i\Gamma^{ }_R\left(f^{ }_R+r^{ }_0f^{ }_L\right)+
             \sqrt{r^{ }_0\Gamma^{ }_L\Gamma^{ }_R} e^{i\varphi_{AB}}
             \left(f^{ }_R-f^{ }_L\right)\right]  .
\end{eqnarray}
\end{widetext}
Therefore $G^{ }_{RL}$ and $G^{ }_{dL}$ may be written in
terms of F's and the full QD Green function as
\begin{eqnarray}
t_Lt_R G^{ }_{RL} &=& F^{ }_{Rd}G^{ }_{dd}F^{ }_{dL} \\
-t^{ }_LG^{ }_{dL} &=& G^{ }_{dd}F^{ }_{dL} .
\end{eqnarray}
Using the Keldysh Green function relations~\cite{Wilkins Langreth}
\begin{eqnarray}
(AB)^< &=& A^<B^a + A^rB^< \\
(ABC)^< &=& A^<B^aC^a + A^rB^<C^a + \nonumber \\
            & &  A^rB^rC^< ,
\end{eqnarray}
the lesser Green functions become
\begin{eqnarray}
G^<_{RL} &=& F^<_{Rd}G^a_{dd}F^{a}_{Ld} + \nonumber \\
             & &  F^{r}_{Rd}G^<_{dd}F^{a}_{Ld} +F^{<}_{Rd}G^a_{dd}F^{a}_{Ld} \\
-t^{ }_LG^<_{dL} &=& G^<_{dd}F^a_{dL} + G^r_{dd}F^<_{dL}  .
\end{eqnarray}
Making use of the nonequilibrium steady state condition,
$\mathrm{i}_L+\mathrm{i}_R = 0$,
we may construct $G^<_{dd}$ in terms of $G^r_{dd}$ and $G^a_{dd}$ as follows.
The ensemble averaged currents are given by
\begin{eqnarray}\label{i_alpha}
\mathrm{i}^{ }_\alpha &=& -t^{ }_\alpha\left(G^<_{d\alpha}-G^<_{\alpha d}\right)  \nonumber\\
             &=& (F^a_{d\alpha}-F^r_{\alpha d})G^<_{dd}
                 +G^r_{dd}F^<_{d\alpha} + F^<_{\alpha d}G^a_{dd}
\end{eqnarray}
where $\alpha=(L,R)$.  Therefore by invoking the steady state condition the QD
lesser Green function becomes
\begin{equation}
G^<_{dd} = \frac{(F^<_{dL}+F^<_{dR})G^r_{dd}-(F^<_{Ld}+F^<_{Rd})G^a_{dd}}
                 {(F^r_{Ld}-F^a_{dL})+(F^r_{Rd}-F^a_{dR})}  .
\end{equation}
Inserting $G^<_{dd}$ into Eq.~(\ref{i_alpha}) with $\alpha=L$ we arrive
at $\mathrm{i}_{QD}$, Eq.~(\ref{i_QD}).  The current through the AB arm
is given by
\begin{widetext}
\begin{eqnarray}
t_Lt_R\mathrm{i}^{ }_{AB} &=& t_Lt_R(-t^*_0G^<_{RL}+t^{ }_0G^<_{LR}) \nonumber \\
                       &=& (-t^*_0F^r_{Rd}F^<_{dL}+t^{ }_0F^r_{Ld}F^<_{dR})G^r_{dd} +
                         (-t^*_0F^<_{Rd}F^a_{dL}+t^{ }_0F^<_{Ld}F^a_{dR})G^a_{dd} + \nonumber \\
                         & &  (-t^*_0F^r_{Rd}F^a_{dL}+t^{ }_0F^r_{Ld}F^a_{dR})G^<_{dd} ,
\end{eqnarray}
\end{widetext}
where all of the terms have been solved for.
Thus after doing some algebra, $\mathrm{i}^{ }_{AB}$
may be expressed as Eq.~(\ref{i_AB}).  Using Eq.~(\ref{T}), we arrive
at the transmission function, Eq.~(\ref{Transmission}).


\begin{thebibliography}{*}
\bibitem{comparison} The percent difference between $G(0)$ [$|t_0|/t_{L/R} = 1.90$]
                     calculated using second-order perturbation theory and the exact
                     numerical renormalization group is $<4\%$ for all $\varphi_{AB}$.  
                     Here the asymmetry is as large as $\delta/\bar{\Gamma} = 0.895$.
\end{thebibliography}
\end{document}